\documentstyle[12pt]{article}

\renewcommand{\baselinestretch}{1.75} 

\newcommand{\la}{\lambda}

\newcommand{\La}{\Lambda}

\newcommand{\be}{\begin{eqnarray}}
\newcommand{\ee}{\end{eqnarray}}

\newcommand{\lra}{\longrightarrow}

\newcommand{\pr}{\partial}

\newcommand{\np}{\newpage}
\newcommand{\hs}{\hspace}
\newcommand{\vs}{\vspace}

\newcommand{\nn}{\nonumber}

\begin{document}

\thispagestyle{empty}

\vs*{-25mm}
\begin{flushright}
SMT00/01\\
\end{flushright}

\begin{center}
{\Large{\bf Seiberg-Witten theory for the asymptotic free rank three tensors 
of $SU(N)$}} \\
\vspace{.2in}

\renewcommand{\baselinestretch}{1}
\small
\normalsize

Henric Rhedin\footnote { henric.rhedin@celsius.se}\\
Celsius Consultants AB\\
Chalmers Teknikpark \\
S-412 88 G\"oteborg\\
Sweden\\

\vspace{.2in}

{\bf{Abstract}} \end{center}
\renewcommand{\baselinestretch}{1.1}
\small
\normalsize
\begin{quotation}
\baselineskip14pt
\noindent 
We here give a first indication that there exists a Seiberg-Witten curve for 
$SU(N)$ Seiberg-Witten theory with matter transforming in the 
totally antisymmetric rank three tensor representation.
We present a derivation of the leading order hyperelliptic 
approximation of a curve for this case. Since we are only 
interested in the asymptotic free theory we are restricted to $N=6,7,8$.  
The derivation is carried out by reversed engineering starting from the known 
form of the prepotential at tree level. We also predict the form of the 
one instanton correction to the prepotential. 
\end{quotation}

\np 

\setcounter{page}{1}

A lot of effort has been spent over the years in Seiberg-Witten theory and all 
asymptotic free cases has been accounted for except $SU(N)$ with matter 
transforming in the totally antisymmetric rank three tensor representation. 
Curves have been derived by generalisations \cite{Everybody} of Seiberg and Wittens 
original work \cite{SeibergWitten}, 
by first principles from M-theory \cite{Mtheory} and most recently by 
reversed engineering \cite{EnnesNaculichRhedinSchnitzer1}. 
Furthermore, results that can be subject to tests of those curves have been 
made available for all cases 
\cite{DHokerKricheverPhong,EnnesNaculichRhedinSchnitzer,EnnesNaculichRhedinSchnitzer1}. 

Here we start a program to determine the Seiberg-Witten curve for $SU(N)$ with matter 
transforming in the totally antisymmetric rank three tensor representation. 
Up to this point no information has been presented for this case and fewer 
clues are available than for e.g. the case of two antisymmetric two tensors 
\cite{EnnesNaculichRhedinSchnitzer1} that was derived by similar methods. 

One of the essential ingredients that characterise the Seiberg-Witten theory is a 
particular residue function denoted $S(x)$ below. For the case studied here there 
is no obvious candidate for this function in contrast to e.g. the 
rank two tensor representations for $SU(N)$. For the rank two tensor representations 
there were also involutions symmetries which from a M-theory perspective implemented 
themselves using known objects such as orientifolds. 
This involution constrained the form of the curve. For the case at hand the realisation 
of such a symmetry at M-theory level is not known and hence the form of the 
curve seems less constrained. 

Never the less one may use the known form of the prepotential at tree level and 
simple arguments like dimension analysis to find a candidate for a 
curve for the $SU(N)$ Seiberg-Witten theory with matter transforming in the 
totally antisymmetric rank three tensor representation. This curve should be 
regarded as the lowest order approximation of a full curve and as a first 
step toward the true answer. 
\\ 

The weights of the totally antisymmetric three tensor representation of SU($N$) 
may be parameterised by 
\be
e_i+e_j+e_k \hs{15 mm}  i< j< k=1,...,N.
\ee
where $e_i$ are the weights of the defining representation. The Dynkin index of the 
three tensor is given by 
\be
I_{Dynkin}=\frac{(N-2)(N-3)}{2}
\ee
and hence asymptotic freedom, which requires $2N-I_{Dynkin}\geq 0$, restrict us to 
$N=6,7$ and 8. We may also add up to 5 and 3 defining representations for $SU(6)$ and 
$SU(7)$ respectively, while $SU(8)$ does not allow for any more matter.   

Take the hyperelliptic curve 
\be
y^2+2A(x)y+L^2B(x)=0, \hs{10 mm} L^2=\La ^{2N-(N-2)(N-3)/2},
\ee 
where $\La$ is the dynamical scale of the theory. 
It is sometimes convenient to use the alternative form 
\be
y^2=(A(x))^2-L^2B(x)
\ee
and we will interchangeably use the notation hyper elliptic approximation for 
either of those two forms. 

The purpose of this paper is to find out whether or not there exists functions 
$A(x)$ and $B(x)$ such that this curve reproduces the tree level prepotential using 
the Seiberg-Witten method. The tree level part of the prepotential is proportional to 
\be
\sum _{i,j=1}^N (e_i-e_j)^2ln\left (\frac{(e_i-e_j)^2}{\La ^2}\right )-
\sum _{i< j< k=1}^N (e_i+e_j+e_k)^2ln\left (\frac{(e_i+e_j+e_k)^2}{\La ^2}\right ).
\label{treelevel}
\ee
and starting from a hyperelliptic curve 
there is a well known prescription \cite{DHokerKricheverPhong} 
how to check this result. Here the objective is to revese this process. 

We expect to be able to find appropriate functions for the 
curve by reverse engineering following the 
method developed in \cite{EnnesNaculichRhedinSchnitzer1}. Reversed engineering makes use 
of symmetries, functional forms, and dimensional restrictions to find a candidate 
for a curve. 
The restriction available apart from 
the form of the tree level prepotential (\ref{treelevel}) is the R-parity of the 
1-instanton 
correction or equivalently the curve. 
Although the form of the 1-instanton contribution to the 
prepotential is not known it will follow from the 
form of $A(x)$ and $B(x)$ and since the R-parity of the 1-instanton term is known 
this puts restrictions on $A(x)$ and $B(x)$. 

When integrating out the matter one should find pure Yang-Mills and hence we 
expect the usual $\prod_{i=1}^N(x-e_i)$ to be a 
part of $A(x)$. The form of the weights indicates 
that $B(x)$ should contain a factor of $\prod (x+e_i+e_j)$. Here there is no obvious choice 
for the range of indices. Options available are $i,j=1,...,N$, $i\leq j=1,...N$ or 
$i<j=1,...N$ and although there is no obvious candidate the minimal choice seems the 
most attractive one for computational reasons. Hence we begin with 
this choice and we will subsequently comment on the other possible choices. 
By inspection it is clear that any of the
choices of $B(x)$ given above will give too many weights and those 
have to be corrected for in some way.

Before we continue we must carry out the calculations to find what kind of 
logarithmic terms follow from the choices 
\be
A(x)=\prod_{i=1}^N(x-e_i) \hs{20 mm} B(x)=\prod_{i<j=1}^N(x+e_i+e_j).
\ee
where we for a moment forget R-parity restrictions. Following Seiberg-Witten 
\cite{SeibergWitten} we would like to calculate the periods $a_i$ and 
dual periods $a_i^D$ 
since the latter are related to the prepotential $F$ via 
\be
a_i^D=\frac{\pr F}{\pr a_i}.
\ee  
The periods and dual periods follow from the usual expression in terms of integrals 
over the cycles $A_k$ and dual cycles $B_k$ as 
\be
a_i=\oint _{A_k} \la dx \hs{20 mm} a_i^D=\oint _{B_k} \la dx \hs{20mm}
\la=\frac{dy}{y}.
\ee

The cycles follow from the shape of the curve and the full set of cycles is 
not known without knowing the exact form of the curve. However, we can find a subset of 
cycles and find their corresponding periods' 
contribution to the tree level prepotential. Additional cycles 
required to meet the constraints on R-parity and correct number of 
weights will give us additional contributions. 

In order to carry out the integrals we need the branchcut structure of the 
curve. In its current form the curve have $N$ branchcuts and the branchpoints follow 
from the constraint that those are the common points of the two sheets $y_+$ and $y_-$ 
where  
\be
y_{\pm}=\sqrt{A^2(x)-L^2B(x)}. 
\ee 
For convenience we introduce the residue function 
\be
S(x)=\frac{B(x)}{A^2(x)} \hs{20 mm} S_k(x)=(x-e_k)^2S(x).
\ee
In terms of this function the $2N$ branchpoints take the form 
\be
x_k^{\pm}=e_k\pm L\sqrt{S_k(e_k)}+...
\ee
and we chose to let the corresponding $N$ branchcuts to run between $x^+_k$ and $x^-_k$. 
We also chose the cycles $A_k$ to surround $x^+_k$ and $x^-_k$ on one sheet and 
the dual cycles $B_k$ to run from $x^+_k$ to $x^-_k$ on one sheet and 
back on the other sheet. 
Note that there may be corrections at this order but since we will only 
be interested in tree level prepotential we only use the zeroth order term 
$x_k^{\pm}=e_k$ although we depend on the existence of a branch cut. 

Standard integration gives the first orders of the periods as 
\be
a_k=e_k+\frac{L^2}{4}\pr_kS_k+...
\ee 
which may also have corrections at this order. We are, however, only keeping 
the lowest order. 
The interesting terms of the dual periods turns out to be 
\be
2\sum_{i=1}^N(e_k-e_i)ln(e_k-e_i)-\sum_{i<j=1}^N(e_k+e_i+e_j)ln(e_k+e_i+e_j)
\ee
which should be compared to the derivative of the prepotential which is proportional to 
\be
2\sum_{i=1}^N(e_k-e_i)ln(e_k-e_i)-\sum_{i<j=1,i,j\neq k}^N(e_k+e_i+e_j)ln(e_k+e_i+e_j).
\ee
The difference is given by terms of the form $(2e_k+e_i)ln(2e_k+e_i)$ 
$i=1,...N,\,\, i\neq k$ which must be corrected by a counterterm of opposite 
sign. It seems obvious to change the residue function into  
\be
S(x)=\frac{\prod_{i<j=1}^N(x+e_i+e_j)}{\prod_{i=1}^N(2x+e_i)\prod_{i=1}^N(x-e_i)^2} 
\label{firstsugg}
\ee
but this does not work for two reasons. It does not correspond to a hyperelliptic 
type of curve since the denominator is not a perfect square. Furthermore, as we will 
show below, there is no way to simultaneously 
get the correct R-parity and weight count using this form. 
Changing $(2x+e_i)$ 
into $(2x+e_i)^2$ gives a perfect square and may give correct R-parity but does not 
provide a correct weight count. There is a 
second possibility that satisfies the hyperelliptic curve demand namely 
\be
S(x)=\frac{\prod_{i<j=1}^N(x+e_i+e_j)}{\prod_{i=1}^N(x+e_i/2)^2\prod_{i=1}^N(x-e_i)^2}.
\ee
Both these forms contribute additional terms to the logarithmic parts of the prepotential 
by 
\be
\sum_{i=1}^N(2e_k+e_i)ln(2e_k+e_i) \label{addons}
\ee
but they have distinct R-parity contributions. Furthermore, there is an over count in 
(\ref{addons}) by $3e_klne_k$ which is corrected for by multiplication by $x^3$. 
We hence suggest the form of the residue function 
\be
S(x)=\frac{x^3\prod_{i<j=1}^N(x+e_i+e_j)}{\prod_{i=1}^N(x-e_i)^2\prod_{i=1}^N(x+e_i/2)^2}.
\label{almostdone}
\ee

The R-parity of $L^2S_k(x)$ should be 2 since we anticipate this form to contribute 
to the 1-instanton correction of the prepotential. Using the R-parity 
of $L^2$ which is $2N-(N-2)(N-3)/2$ 
we find the form above (\ref{almostdone}) to have the correct R-parity. Note that 
the first suggestion (\ref{firstsugg}) would give R-parity of $L^2S_k(x)$ to be $2+N-3$. 
We could have introduced an additional $x^{3-N}$ to ensure the correct R-parity but 
this would ruin the weight count. 

We now have a new form of curve with functions 
\be
A(x)=\prod_{i=1}^N(x+e_i/2)\prod_{i=1}^N(x-e_i), \hs{20mm} 
B(x)=x^3\prod_{i<j=1}^N(x+e_i+e_j).\label{almosttrue}
\ee
This suggestion has, however, a serious flaw. As can be seen from the branchpoints 
discussion above this set of functions (\ref{almosttrue}) forces us to consider 
more branchcuts centred on $x=-e_i/2$. This also means more cycles which will 
yield additional contributions to the prepotential. A closer study of the curve 
for the antisymmetric two tensor representation indicates a way out. If we multiply 
$A(x)$ by another factor of $\prod_{i=1}^N(x+e_i/2)$ and 
$B(x)$ by $\prod_{i=1}^N(x+e_i/2)^2$ then the ramification points where the 
two sheets coincide will not require branchcuts and hence there are no 
additional cycles. Furthermore, this does not change R-parity of the 
residue function $S(x)$.

The final form of the hyperelliptic approximation is thus 
\be
y^2+2A(x)y+L^2B(x)=0 & & \hs{5mm}{\rm where}\\
A(x)=\prod_{i=1}^N(x+e_i/2)^2\prod_{i=1}^N(x-e_i)& & \hs{5mm} 
B(x)=x^3\prod_{i=1}^N(x+e_i/2)^2\prod_{i<j=1}^N(x+e_i+e_j).\label{done}\nn
\ee
This curve gives the correct 1-loop prepotential for $SU(N)$ with matter in the 
totally antisymmetric three tensor representation. It is also clear that by introducing 
a mass $m$ for the matter representation  in the usual way by taking $x\lra x+m$ 
and $e_i\lra e_i+m$ one may as usual integrate out $m$ yielding the correct 
pure Yang-Mills result. 

Defining matter may be incorporated by multiplication of additional factors
\be
\prod_{i=1}^{N_f}(x+M_i)
\ee
to $B(x)$. Here $N_f\leq 5$ for $N=6$ and $N_f\leq 3$ for $N=7$ while $N=8$ 
does not allow for any defining matter. 

We mentioned above the possibility to have another $B(x)$ with a wider range of 
indices e.g. $i\leq j=1,...N$ or $i,j=1,...N$ instead of $i< j=1,...N$ as above. 
We have not been able 
to find a form of $S(x)$ that would yield the correct number of weights and at the 
same time respect R-parity restrictions for any other choice of range of indices.  

We now proceed by predicting the one instanton contribution to the prepotential. 
Following the same line of reasoning as in \cite{EnnesNaculichRhedinSchnitzer1} we 
give that the following form of the one instanton contribution 
\be
F_{1-inst}=\sum_{k=1}^N(S_k(e_k)-2\bar{S}_k(-e_k/2-3m/2))
\ee
where 
\be
S(x)=\frac{(x+m)^3\prod_{i<j=1}^N(x+e_i+e_j+3m)\prod_{j=1}^{N_f}(x+M_j)}{
\prod_{i=1}^N(x+e_i/2+3m/2)^2\prod_{i=1}^N(x-e_i)^2}
\hs{10 mm} && \rm{and} \\
S_k(x)\equiv (x-e_k)^2S(x), \hs{10 mm} 
\bar{S}_k(x)\equiv (x+e_k/2+3m/2)^2S(x). && \nn
\ee
This result has the following properties. It yields the correct pure Yang-Mills 
and defining flavour results in the double scaling limit $m\rightarrow \infty$ and 
$L^2m^{3+n(n-1)/2-2N}\rightarrow L^2_{new}$ where $L^2_{new}$ is the new scale of 
the theory. Furtermore, it does not have any poles as $e_k=-e_l/2-3m/2$ for 
some $l\in\{1,...N\}$, which is the fact for 
a sum over $S_k(e_k)$ only. The precence of the sum over $\bar{S}_k(e_k)$ guarantees 
that there are only poles in the appropriate places. 

It is clear that this curve is not the full curve. It is expected to have subleading 
(contributions with $L^2$ or higher order) terms in $A(x)$ as was the case for the 
antisymmetric rank two tensor. Also, the curve itself is expected to have a higher 
degree. We hope to be able to present more details on those issues in the near future.

\end{document}